\documentclass[twocolumn,final]{svjour3}

\AtBeginDocument{
}

\usepackage{mathptmx}
\usepackage{upgreek}
\usepackage[numbers,super]{natbib}

\usepackage{caption}
\DeclareCaptionLabelFormat{default}{}

\makeatletter
\def\longtable{\@ifnextchar[\newlongtable@i \newlongtable@ii}
\def\newlongtable@i[#1]{%
\renewcommand{\endhead}{\ignorespaces}
\xtabular[#1]}
\def\newlongtable@ii{%
\renewcommand{\endhead}{\ignorespaces}
\xtabular}

\makeatother

    \usepackage[breakable]{tcolorbox}
    \usepackage{parskip} 
    
    \usepackage{iftex}
    \ifPDFTeX
    	\usepackage[T1]{fontenc}
    	\usepackage{mathpazo}
    \else
    	\usepackage{fontspec}
    \fi

    \usepackage{graphicx}
    
    \usepackage{caption}

    \usepackage[Export]{adjustbox} 
    \adjustboxset{max size={0.9\linewidth}{0.9\paperheight}}
    \usepackage{float}
    \usepackage{xcolor} 
    \usepackage{enumerate} 
    \usepackage{geometry} 
    \usepackage{amsmath} 
    \usepackage{amssymb} 
    \usepackage{textcomp} 
    \AtBeginDocument{%
    }
    \usepackage{upquote} 
    \usepackage{eurosym} 
    \usepackage[mathletters]{ucs} 
    \usepackage{fancyvrb} 
    \usepackage{grffile} 
    \makeatletter 
    \def\Gread@@xetex#1{%
      \IfFileExists{"\Gin@base".bb}%
      {\Gread@eps{\Gin@base.bb}}%
      {\Gread@@xetex@aux#1}%
    }
    \makeatother

    \usepackage{hyperref}
    
    \usepackage{longtable} 
    \usepackage{booktabs}  
    \usepackage[inline]{enumitem} 
    \usepackage[normalem]{ulem} 
    \usepackage{mathrsfs}

    \definecolor{urlcolor}{rgb}{0,.145,.698}
    \definecolor{linkcolor}{rgb}{.71,0.21,0.01}
    \definecolor{citecolor}{rgb}{.12,.54,.11}

    \definecolor{ansi-black}{HTML}{3E424D}
    \definecolor{ansi-black-intense}{HTML}{282C36}
    \definecolor{ansi-red}{HTML}{E75C58}
    \definecolor{ansi-red-intense}{HTML}{B22B31}
    \definecolor{ansi-green}{HTML}{00A250}
    \definecolor{ansi-green-intense}{HTML}{007427}
    \definecolor{ansi-yellow}{HTML}{DDB62B}
    \definecolor{ansi-yellow-intense}{HTML}{B27D12}
    \definecolor{ansi-blue}{HTML}{208FFB}
    \definecolor{ansi-blue-intense}{HTML}{0065CA}
    \definecolor{ansi-magenta}{HTML}{D160C4}
    \definecolor{ansi-magenta-intense}{HTML}{A03196}
    \definecolor{ansi-cyan}{HTML}{60C6C8}
    \definecolor{ansi-cyan-intense}{HTML}{258F8F}
    \definecolor{ansi-white}{HTML}{C5C1B4}
    \definecolor{ansi-white-intense}{HTML}{A1A6B2}
    \definecolor{ansi-default-inverse-fg}{HTML}{FFFFFF}
    \definecolor{ansi-default-inverse-bg}{HTML}{000000}

    \providecommand{\tightlist}{%
      \setlength{\itemsep}{0pt}\setlength{\parskip}{0pt}}
    \DefineVerbatimEnvironment{Highlighting}{Verbatim}{commandchars=\\\{\}}

    \let\Oldtex\TeX
    \let\Oldlatex\LaTeX
    \renewcommand{\TeX}{\textrm{\Oldtex}}
    \renewcommand{\LaTeX}{\textrm{\Oldlatex}}
    \title{Arxiv}

\makeatletter
\def\PY@reset{\let\PY@it=\relax \let\PY@bf=\relax%
    \let\PY@ul=\relax \let\PY@tc=\relax%
    \let\PY@bc=\relax \let\PY@ff=\relax}
\def\PY@tok#1{\csname PY@tok@#1\endcsname}
\def\PY@toks#1+{\ifx\relax#1\empty\else%
    \PY@tok{#1}\expandafter\PY@toks\fi}
\def\PY@do#1{\PY@bc{\PY@tc{\PY@ul{%
    \PY@it{\PY@bf{\PY@ff{#1}}}}}}}
\def\PY#1#2{\PY@reset\PY@toks#1+\relax+\PY@do{#2}}

\expandafter\def\csname PY@tok@w\endcsname{\def\PY@tc##1{\textcolor[rgb]{0.73,0.73,0.73}{##1}}}
\expandafter\def\csname PY@tok@c\endcsname{\let\PY@it=\textit\def\PY@tc##1{\textcolor[rgb]{0.25,0.50,0.50}{##1}}}
\expandafter\def\csname PY@tok@cp\endcsname{\def\PY@tc##1{\textcolor[rgb]{0.74,0.48,0.00}{##1}}}
\expandafter\def\csname PY@tok@k\endcsname{\let\PY@bf=\textbf\def\PY@tc##1{\textcolor[rgb]{0.00,0.50,0.00}{##1}}}
\expandafter\def\csname PY@tok@kp\endcsname{\def\PY@tc##1{\textcolor[rgb]{0.00,0.50,0.00}{##1}}}
\expandafter\def\csname PY@tok@kt\endcsname{\def\PY@tc##1{\textcolor[rgb]{0.69,0.00,0.25}{##1}}}
\expandafter\def\csname PY@tok@o\endcsname{\def\PY@tc##1{\textcolor[rgb]{0.40,0.40,0.40}{##1}}}
\expandafter\def\csname PY@tok@ow\endcsname{\let\PY@bf=\textbf\def\PY@tc##1{\textcolor[rgb]{0.67,0.13,1.00}{##1}}}
\expandafter\def\csname PY@tok@nb\endcsname{\def\PY@tc##1{\textcolor[rgb]{0.00,0.50,0.00}{##1}}}
\expandafter\def\csname PY@tok@nf\endcsname{\def\PY@tc##1{\textcolor[rgb]{0.00,0.00,1.00}{##1}}}
\expandafter\def\csname PY@tok@nc\endcsname{\let\PY@bf=\textbf\def\PY@tc##1{\textcolor[rgb]{0.00,0.00,1.00}{##1}}}
\expandafter\def\csname PY@tok@nn\endcsname{\let\PY@bf=\textbf\def\PY@tc##1{\textcolor[rgb]{0.00,0.00,1.00}{##1}}}
\expandafter\def\csname PY@tok@ne\endcsname{\let\PY@bf=\textbf\def\PY@tc##1{\textcolor[rgb]{0.82,0.25,0.23}{##1}}}
\expandafter\def\csname PY@tok@nv\endcsname{\def\PY@tc##1{\textcolor[rgb]{0.10,0.09,0.49}{##1}}}
\expandafter\def\csname PY@tok@no\endcsname{\def\PY@tc##1{\textcolor[rgb]{0.53,0.00,0.00}{##1}}}
\expandafter\def\csname PY@tok@nl\endcsname{\def\PY@tc##1{\textcolor[rgb]{0.63,0.63,0.00}{##1}}}
\expandafter\def\csname PY@tok@ni\endcsname{\let\PY@bf=\textbf\def\PY@tc##1{\textcolor[rgb]{0.60,0.60,0.60}{##1}}}
\expandafter\def\csname PY@tok@na\endcsname{\def\PY@tc##1{\textcolor[rgb]{0.49,0.56,0.16}{##1}}}
\expandafter\def\csname PY@tok@nt\endcsname{\let\PY@bf=\textbf\def\PY@tc##1{\textcolor[rgb]{0.00,0.50,0.00}{##1}}}
\expandafter\def\csname PY@tok@nd\endcsname{\def\PY@tc##1{\textcolor[rgb]{0.67,0.13,1.00}{##1}}}
\expandafter\def\csname PY@tok@s\endcsname{\def\PY@tc##1{\textcolor[rgb]{0.73,0.13,0.13}{##1}}}
\expandafter\def\csname PY@tok@sd\endcsname{\let\PY@it=\textit\def\PY@tc##1{\textcolor[rgb]{0.73,0.13,0.13}{##1}}}
\expandafter\def\csname PY@tok@si\endcsname{\let\PY@bf=\textbf\def\PY@tc##1{\textcolor[rgb]{0.73,0.40,0.53}{##1}}}
\expandafter\def\csname PY@tok@se\endcsname{\let\PY@bf=\textbf\def\PY@tc##1{\textcolor[rgb]{0.73,0.40,0.13}{##1}}}
\expandafter\def\csname PY@tok@sr\endcsname{\def\PY@tc##1{\textcolor[rgb]{0.73,0.40,0.53}{##1}}}
\expandafter\def\csname PY@tok@ss\endcsname{\def\PY@tc##1{\textcolor[rgb]{0.10,0.09,0.49}{##1}}}
\expandafter\def\csname PY@tok@sx\endcsname{\def\PY@tc##1{\textcolor[rgb]{0.00,0.50,0.00}{##1}}}
\expandafter\def\csname PY@tok@m\endcsname{\def\PY@tc##1{\textcolor[rgb]{0.40,0.40,0.40}{##1}}}
\expandafter\def\csname PY@tok@gh\endcsname{\let\PY@bf=\textbf\def\PY@tc##1{\textcolor[rgb]{0.00,0.00,0.50}{##1}}}
\expandafter\def\csname PY@tok@gu\endcsname{\let\PY@bf=\textbf\def\PY@tc##1{\textcolor[rgb]{0.50,0.00,0.50}{##1}}}
\expandafter\def\csname PY@tok@gd\endcsname{\def\PY@tc##1{\textcolor[rgb]{0.63,0.00,0.00}{##1}}}
\expandafter\def\csname PY@tok@gi\endcsname{\def\PY@tc##1{\textcolor[rgb]{0.00,0.63,0.00}{##1}}}
\expandafter\def\csname PY@tok@gr\endcsname{\def\PY@tc##1{\textcolor[rgb]{1.00,0.00,0.00}{##1}}}
\expandafter\def\csname PY@tok@ge\endcsname{\let\PY@it=\textit}
\expandafter\def\csname PY@tok@gs\endcsname{\let\PY@bf=\textbf}
\expandafter\def\csname PY@tok@gp\endcsname{\let\PY@bf=\textbf\def\PY@tc##1{\textcolor[rgb]{0.00,0.00,0.50}{##1}}}
\expandafter\def\csname PY@tok@go\endcsname{\def\PY@tc##1{\textcolor[rgb]{0.53,0.53,0.53}{##1}}}
\expandafter\def\csname PY@tok@gt\endcsname{\def\PY@tc##1{\textcolor[rgb]{0.00,0.27,0.87}{##1}}}
\expandafter\def\csname PY@tok@err\endcsname{\def\PY@bc##1{\setlength{\fboxsep}{0pt}\fcolorbox[rgb]{1.00,0.00,0.00}{1,1,1}{\strut ##1}}}
\expandafter\def\csname PY@tok@kc\endcsname{\let\PY@bf=\textbf\def\PY@tc##1{\textcolor[rgb]{0.00,0.50,0.00}{##1}}}
\expandafter\def\csname PY@tok@kd\endcsname{\let\PY@bf=\textbf\def\PY@tc##1{\textcolor[rgb]{0.00,0.50,0.00}{##1}}}
\expandafter\def\csname PY@tok@kn\endcsname{\let\PY@bf=\textbf\def\PY@tc##1{\textcolor[rgb]{0.00,0.50,0.00}{##1}}}
\expandafter\def\csname PY@tok@kr\endcsname{\let\PY@bf=\textbf\def\PY@tc##1{\textcolor[rgb]{0.00,0.50,0.00}{##1}}}
\expandafter\def\csname PY@tok@bp\endcsname{\def\PY@tc##1{\textcolor[rgb]{0.00,0.50,0.00}{##1}}}
\expandafter\def\csname PY@tok@fm\endcsname{\def\PY@tc##1{\textcolor[rgb]{0.00,0.00,1.00}{##1}}}
\expandafter\def\csname PY@tok@vc\endcsname{\def\PY@tc##1{\textcolor[rgb]{0.10,0.09,0.49}{##1}}}
\expandafter\def\csname PY@tok@vg\endcsname{\def\PY@tc##1{\textcolor[rgb]{0.10,0.09,0.49}{##1}}}
\expandafter\def\csname PY@tok@vi\endcsname{\def\PY@tc##1{\textcolor[rgb]{0.10,0.09,0.49}{##1}}}
\expandafter\def\csname PY@tok@vm\endcsname{\def\PY@tc##1{\textcolor[rgb]{0.10,0.09,0.49}{##1}}}
\expandafter\def\csname PY@tok@sa\endcsname{\def\PY@tc##1{\textcolor[rgb]{0.73,0.13,0.13}{##1}}}
\expandafter\def\csname PY@tok@sb\endcsname{\def\PY@tc##1{\textcolor[rgb]{0.73,0.13,0.13}{##1}}}
\expandafter\def\csname PY@tok@sc\endcsname{\def\PY@tc##1{\textcolor[rgb]{0.73,0.13,0.13}{##1}}}
\expandafter\def\csname PY@tok@dl\endcsname{\def\PY@tc##1{\textcolor[rgb]{0.73,0.13,0.13}{##1}}}
\expandafter\def\csname PY@tok@s2\endcsname{\def\PY@tc##1{\textcolor[rgb]{0.73,0.13,0.13}{##1}}}
\expandafter\def\csname PY@tok@sh\endcsname{\def\PY@tc##1{\textcolor[rgb]{0.73,0.13,0.13}{##1}}}
\expandafter\def\csname PY@tok@s1\endcsname{\def\PY@tc##1{\textcolor[rgb]{0.73,0.13,0.13}{##1}}}
\expandafter\def\csname PY@tok@mb\endcsname{\def\PY@tc##1{\textcolor[rgb]{0.40,0.40,0.40}{##1}}}
\expandafter\def\csname PY@tok@mf\endcsname{\def\PY@tc##1{\textcolor[rgb]{0.40,0.40,0.40}{##1}}}
\expandafter\def\csname PY@tok@mh\endcsname{\def\PY@tc##1{\textcolor[rgb]{0.40,0.40,0.40}{##1}}}
\expandafter\def\csname PY@tok@mi\endcsname{\def\PY@tc##1{\textcolor[rgb]{0.40,0.40,0.40}{##1}}}
\expandafter\def\csname PY@tok@il\endcsname{\def\PY@tc##1{\textcolor[rgb]{0.40,0.40,0.40}{##1}}}
\expandafter\def\csname PY@tok@mo\endcsname{\def\PY@tc##1{\textcolor[rgb]{0.40,0.40,0.40}{##1}}}
\expandafter\def\csname PY@tok@ch\endcsname{\let\PY@it=\textit\def\PY@tc##1{\textcolor[rgb]{0.25,0.50,0.50}{##1}}}
\expandafter\def\csname PY@tok@cm\endcsname{\let\PY@it=\textit\def\PY@tc##1{\textcolor[rgb]{0.25,0.50,0.50}{##1}}}
\expandafter\def\csname PY@tok@cpf\endcsname{\let\PY@it=\textit\def\PY@tc##1{\textcolor[rgb]{0.25,0.50,0.50}{##1}}}
\expandafter\def\csname PY@tok@c1\endcsname{\let\PY@it=\textit\def\PY@tc##1{\textcolor[rgb]{0.25,0.50,0.50}{##1}}}
\expandafter\def\csname PY@tok@cs\endcsname{\let\PY@it=\textit\def\PY@tc##1{\textcolor[rgb]{0.25,0.50,0.50}{##1}}}

\makeatother

    \makeatletter
        \newbox\Wrappedcontinuationbox 
        \newbox\Wrappedvisiblespacebox 
        \newcommand*\Wrappedvisiblespace {\textcolor{red}{\textvisiblespace}} 
        \newcommand*\Wrappedcontinuationsymbol {\textcolor{red}{\llap{\tiny$\m@th\hookrightarrow$}}} 
        \newcommand*\Wrappedcontinuationindent {3ex } 
        \newcommand*\Wrappedafterbreak {\kern\Wrappedcontinuationindent\copy\Wrappedcontinuationbox} 
        \newcommand*\Wrappedbreaksatspecials {%
            \def\PYGZus{\discretionary{\char`\_}{\Wrappedafterbreak}{\char`\_}}%
            \def\PYGZob{\discretionary{}{\Wrappedafterbreak\char`\{}{\char`\{}}%
            \def\PYGZcb{\discretionary{\char`\}}{\Wrappedafterbreak}{\char`\}}}%
            \def\PYGZca{\discretionary{\char`\^}{\Wrappedafterbreak}{\char`\^}}%
            \def\PYGZam{\discretionary{\char`\&}{\Wrappedafterbreak}{\char`\&}}%
            \def\PYGZlt{\discretionary{}{\Wrappedafterbreak\char`\<}{\char`\<}}%
            \def\PYGZgt{\discretionary{\char`\>}{\Wrappedafterbreak}{\char`\>}}%
            \def\PYGZsh{\discretionary{}{\Wrappedafterbreak\char`\#}{\char`\#}}%
            \def\PYGZpc{\discretionary{}{\Wrappedafterbreak\char`\%}{\char`\%}}%
            \def\PYGZdl{\discretionary{}{\Wrappedafterbreak\char`\$}{\char`\$}}%
            \def\PYGZhy{\discretionary{\char`\-}{\Wrappedafterbreak}{\char`\-}}%
            \def\PYGZsq{\discretionary{}{\Wrappedafterbreak\textquotesingle}{\textquotesingle}}%
            \def\PYGZdq{\discretionary{}{\Wrappedafterbreak\char`\"}{\char`\"}}%
            \def\PYGZti{\discretionary{\char`\~}{\Wrappedafterbreak}{\char`\~}}%
        } 
        \newcommand*\Wrappedbreaksatpunct {%
            \lccode`\~`\.\lowercase{\def~}{\discretionary{\hbox{\char`\.}}{\Wrappedafterbreak}{\hbox{\char`\.}}}%
            \lccode`\~`\,\lowercase{\def~}{\discretionary{\hbox{\char`\,}}{\Wrappedafterbreak}{\hbox{\char`\,}}}%
            \lccode`\~`\;\lowercase{\def~}{\discretionary{\hbox{\char`\;}}{\Wrappedafterbreak}{\hbox{\char`\;}}}%
            \lccode`\~`\:\lowercase{\def~}{\discretionary{\hbox{\char`\:}}{\Wrappedafterbreak}{\hbox{\char`\:}}}%
            \lccode`\~`\?\lowercase{\def~}{\discretionary{\hbox{\char`\?}}{\Wrappedafterbreak}{\hbox{\char`\?}}}%
            \lccode`\~`\!\lowercase{\def~}{\discretionary{\hbox{\char`\!}}{\Wrappedafterbreak}{\hbox{\char`\!}}}%
            \lccode`\~`\/\lowercase{\def~}{\discretionary{\hbox{\char`\/}}{\Wrappedafterbreak}{\hbox{\char`\/}}}%
            \catcode`\.\active
            \catcode`\,\active 
            \catcode`\;\active
            \catcode`\:\active
            \catcode`\?\active
            \catcode`\!\active
            \catcode`\/\active 
            \lccode`\~`\~ 	
        }
    \makeatother

    \let\OriginalVerbatim=\Verbatim
    \makeatletter
    \renewcommand{\Verbatim}[1][1]{%
        \sbox\Wrappedcontinuationbox {\Wrappedcontinuationsymbol}%
        \sbox\Wrappedvisiblespacebox {\FV@SetupFont\Wrappedvisiblespace}%
        \def\FancyVerbFormatLine ##1{\hsize\linewidth
            \vtop{\raggedright\hyphenpenalty\z@\exhyphenpenalty\z@
                \doublehyphendemerits\z@\finalhyphendemerits\z@
                \strut ##1\strut}%
        }%
        \def\FV@Space {%
            \nobreak\hskip\z@ plus\fontdimen3\font minus\fontdimen4\font
            \discretionary{\copy\Wrappedvisiblespacebox}{\Wrappedafterbreak}
            {\kern\fontdimen2\font}%
        }%
        
        \Wrappedbreaksatspecials
        \OriginalVerbatim[#1,codes*=\Wrappedbreaksatpunct]%
    }
    \makeatother

    \definecolor{incolor}{HTML}{303F9F}
    \definecolor{outcolor}{HTML}{D84315}
    \definecolor{cellborder}{HTML}{CFCFCF}
    \definecolor{cellbackground}{HTML}{F7F7F7}
    
    \makeatletter
    \newcommand{\boxspacing}{\kern\kvtcb@left@rule\kern\kvtcb@boxsep}
    \makeatother

    \hypersetup{
      breaklinks=true,  
      colorlinks=true,
      urlcolor=urlcolor,
      linkcolor=linkcolor,
      citecolor=citecolor,
      }
    
\begin{document}
    
    \author{Torsten Schenkel \and Ian Halliday}\institute{\email{t.schenkel@shu.ac.uk} \\ Department of Engineering and Mathematics \\ Materials and Engineering Research Institute (MERI) \\ Sheffield Hallam University}\title{Continuum Scale Non Newtonian Particle Transport Model for H{\ae}morheology - Implementation and Validation}\titlerunning{Non Newtonian Particle Transport Model for H{\ae}morheology}

\date{\today}
\maketitle

    \begin{abstract}
    We present a continuum scale particle transport model for red blood
cells following collision arguments in a diffusive flux formulation. The
model is implemented in FOAM, in a framework for haemodynamics
simulations. Modern mechanistic rheology models are implemented and
tested. The model is verified against a known analytical solution and
shows excellent agreement for high quality meshes and good agreement for
typical meshes as used in vascular flow simulations. Simulation results
for different size and time scales show that migration of red blood
cells does occur on physiologically relevany timescales on small vessels
below \(1\ mm\) and that the haematocrit concentration modulates the
non-Newtonian viscosity. This model forms part of a multi-scale approach
to haemorheology and model parameters will be derived from meso-scale
simulations using multi-component Lattice-Boltzmann methods. The code,
\texttt{haemoFoam}, is made available for interested researchers.
\end{abstract}
    \hypertarget{introduction}{%
\section{Introduction}\label{introduction}}

Blood is a non-Newtonian fluid with very complex behaviour deriving from
a mesoscopic composition which - minimally described - is a dense,
mono-disperse suspension of deformable vesicles suspended in
incompressible plasma. Accordingly, blood rheology is dominated by the
interaction of cells, with a multitude of models having been proposed to
account for such meso-scale effects as deformation, aggregation, and
rouleaux formation which underline emergent macroscopic flow properties
like concentration dependant viscosity and shear thinning. The authors
are currently developing a multi-scale approach, explicitly modelling
meso-scale effects using Lattice Boltzmann Models (LBM), in which
erythrocyte mechanics are fully resolved, while describing the
macro-scale rheology using particle transport modelling and
quasi-mechanistic non- Newtonian rheology models. The latter will
eventually be parameterised using LBM data. Here, we present the
continuum mechanical part of the modelling approach, which allows the
simulation of realistic vessel geometries and complex flow patterns.

    \hypertarget{methods}{%
\section{Methods}\label{methods}}

    \hypertarget{particle-migration-model}{%
\subsection{Particle Migration Model}\label{particle-migration-model}}

In a high particle load suspension like blood, many types of mechanical
interactions between particles and carrier fluid take place. Mesoscale
modelling, using the multi-component Lattice-Boltzmann Method
\cite{Aidun2010}, which has widely acknowledged facility for Lagrangian
particulate flows
\cite{Clausen2010,Ladd1994,Ladd1994a,MacMeccan2007,Dupin2007} is
employed to describe these interactions and the dynamics of the
collision in detail.

As with direct numerical simulation in turbulence modelling, finite
computational resource means that detailed explicit particulate models
are limited to small volumes containing relatively few particles in
their simulation domain (an the order of magnitude of hundreds to
thousands at the time of writing). To address the much greater scales of
medical significance, it is, therefore, necessary to develop macro- or
continuum scale models, encapsulating the integral effect of these
interactions without explicitely resolving them. Crucially, these models
must be amenable to parameterisation using meso-scale data, such as
\cite{Burgin2018}. Currently, the models which have been proposed for
this task can, roughly, be divided \cite{Fang2002} into suspension
balance models \cite{Leighton1987,Nott1994} and diffusive flux models
\cite{Phillips1992}.

Suspension balance models use an Euler-Euler mixture modelling approach,
where the carrier fluid and the particle load are represented as
separate species with a transport equation (typically
convection-diffusion) and physical transport properties for each
species, while in the diffusive flux models, the suspension is modelled
as a single species with the particle volume fraction being modelled as
a scalar property, which influences the bulk transport properties.

Our macroscopic model is a particle transport model after Phillips
\cite{Phillips1992} and follows the collision arguments by Leighton and
Acrivios \cite{Leighton1987}. It describes the particle migration based
on the gradients of shear strain, concentration and viscosity. The local
concentration of haematocrit is then used to establish the local
effective viscosity.

A detailed treatise on the rationale behind the compression arguments
can be found in Leighton and Acrvios, and Phillips
\cite{Leighton1987,Phillips1992}, we only give a brief outline at this
point.

The transport of haematocrit is dominated by advection - following the
bulk flow - variations in concentration are evened by diffusive
processes, and the migration within the bulk is driven by a migration
pressure. This migration pressure is the result of two phenomena: (1)
spatial variation of collision (interaction) frequency, and (2) spatial
variation of viscosity.

    \hypertarget{spatial-variation-of-collision-frequency}{%
\subsubsection{Spatial variation of collision
frequency}\label{spatial-variation-of-collision-frequency}}

Particles that are moving relative to each other in neighboring shear
surfaces will experience collisions. The frequency of these collisions
is proportional to the shear rate \(\dot{\gamma}\), the particle
concentration \(\phi\), and the particle collision radius \(a\). In a
field of constant concentration and constant shear,
\(\dot{\gamma}\phi = \text{const}\), the collisions are in equilibrium
either side of the shear surface, and no net migration will occur. In
the presence of gradients of shear rate or concentration, the imbalance
of collisions will lead to a ``migration pressure'' down the gradient.
This collsion driven migration pressure can be described as a function
of \(a \nabla (\dot{\gamma}\phi))\). Using a proportinality factor of
\(K_c\) and assuming a displacement proportional to the particle radius
\(a\), the migratory flux \(N_c\) due to variations in collision
frequency can be expressed as (using the chain rule):

\begin{equation}
    N_c = - K_c a^2 (\phi^2 \nabla \dot{\gamma} + \phi \dot{\gamma} \nabla \phi)
\end{equation}

    \hypertarget{spatial-variation-of-viscosity}{%
\subsubsection{Spatial variation of
viscosity}\label{spatial-variation-of-viscosity}}

The displacement of particles after a collision is moderated by viscous
effects. In a constant viscosity field the displacement is isotropic and
thus balanced with no net migration effects. In a viscosity gradient,
the displacement will be less damped in direction of the lower
viscosity, leading to a net migration effect down the viscisity
gradient.

The displacement velocity is proportional to the relative change in
viscosity over a distance that is of order \(a\):
\(a (1/\mu) \nabla \mu\). With the displacement frequency scaling with
\(\dot{\gamma}\phi\), and a proportionality factor of \(K_\mu\), the
migratory flux due to viscosity gradient can be described as (flux is
proportional to \(\phi\)):

\begin{equation}
    N_\mu = - K_\mu \dot{\gamma}\phi^2 \left( \frac{a^2}{\mu} \nabla \mu \right)
\end{equation}

    The scalar transport equation for haematocrit, \(\phi\), is then
(neglecting molecular diffusion, Brownian motion), where \(D/Dt\) is the
total differential:

\begin{eqnarray}
\frac{D \phi}{D t} = & &\nabla \cdot \left(N_c + N_\mu \right)\\
& & \nonumber \\
\frac{D \phi}{D t} = & &\nabla \cdot \left( a^2 K_c \phi \dot{\gamma} \nabla \phi \right) \nonumber \\
& + & a^2 K_c \nabla \cdot \left( \phi^2 \nabla \dot{\gamma} \right) \nonumber\\
& + & a^2 K_\mu \nabla \cdot \left( \dot{\gamma} \phi^2 \frac{1}{\mu} \nabla \mu \right), \label{eq:phiTransport}
\end{eqnarray}

with \(a\), particle radius, \(\dot{\gamma}\), shear strain rate
magnitude, \(\mu\), dynamic viscosity, \(K_c\) and \(K_\mu\), collision
parameters.

Typically, the viscosity is \(\mu = f(\dot{\gamma},\phi)\), which makes
the last source term non-linear, which can, in turn, make the solution
of this transport equation difficult.

Previous attempts to solve this problem analytically or implement this
type of migration model in a numerical model used linearisation of this
source term, which involves the derivative of \(\mu\) in both
\(\dot{\gamma}\) and \(\phi\), and thus limits the model to a specific
viscosity model, for which it has been implemented
\cite{Mansour2010,Chebbi2018}. Our current implementation deals with the
non-linear viscosity source term in a way that leaves the viscosity
gradient term intact and is thus agnostic to the rheology model used.

    \hypertarget{rheology-models}{%
\subsection{Rheology Models}\label{rheology-models}}

It is obvious from the third RHS term in equation \ref{eq:phiTransport},
that the particle transport strongly depends on the rheology model it is
coupled with. This model implementation aims to be independent of the
rheology model. The draw-back of this approach is that errors present in
the rheology model, which influence the particle transport, cannot be
calibrated out with the parameters of the migration model alone, but the
combined set of model parameters will need to be found for any new
rheology model that is to be implemented.

Typically, only the shear thinning effects are taken into account, when
modelling the non-Newtonian properties of blood in CFD. Common models
are of the Carreau and Casson types (REF). In these models, the
haematocrit concentration is only used as a bulk parameter in the
parametrisation, if at all. Our framework, incorporating the transport
of haematocrit, allows the rheological model to take the local particle
concentration into account when calculating the local, effective
viscosity.

\begin{figure}
            \begin{center}\adjustimage{max size={\linewidth}{0.7\paperheight}}{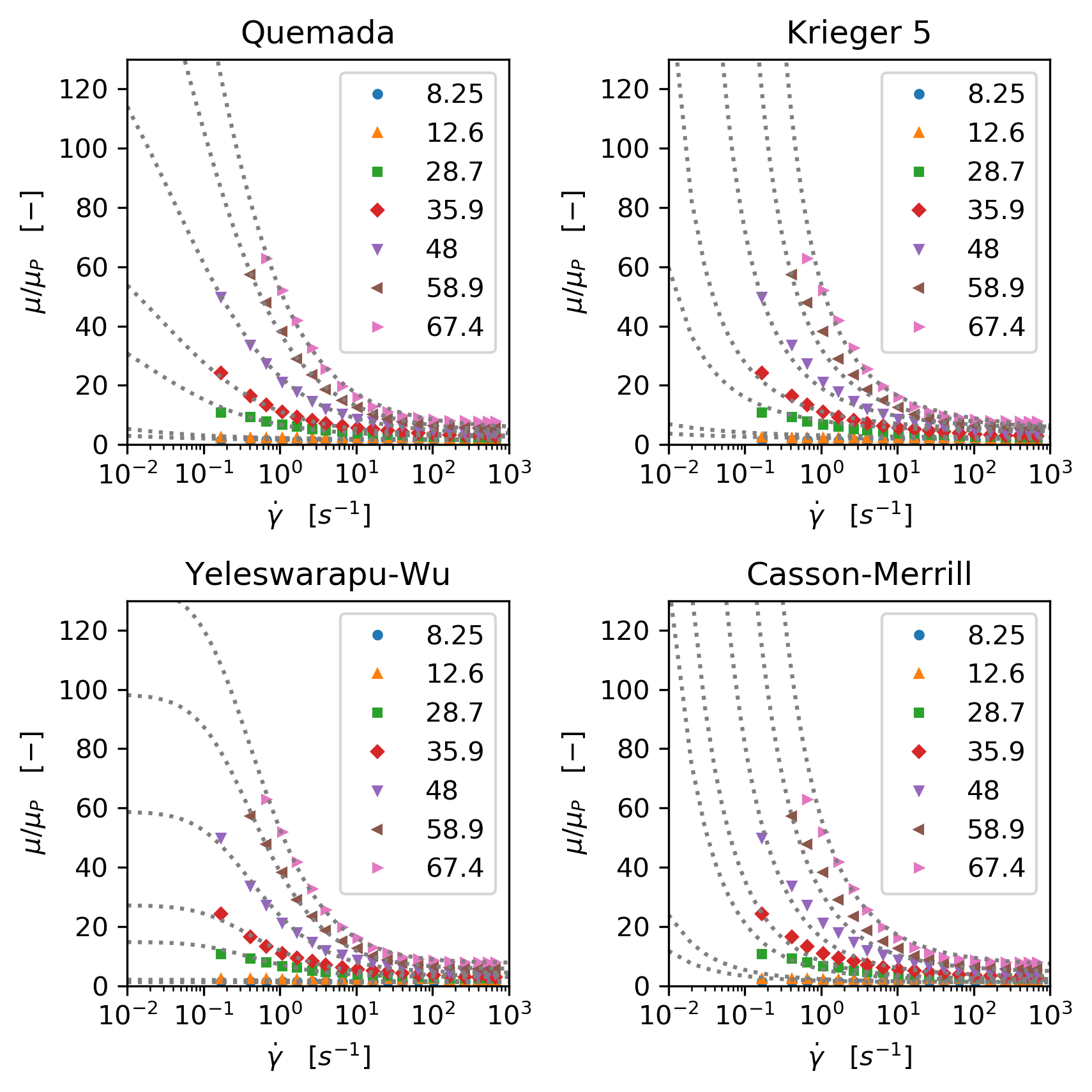}\end{center}
            \caption{Comparison of non-Newtonian rheology models. All model parameters have been fitted to Brooks' data. Dots: experimental data (Brooks), dotted lines: model equations}
            \label{fig:viscosityModels}
        \end{figure}

The rheology models that have been implemented and tested in this study
are the concentration dependent Krieger-Dougherty model
\cite{Krieger1959}, the Quemada model
\cite{Quemada1977,Quemada1978,Quemada1978a} with modification by Das
\cite{Das1998} (and a new parameter set, which avoids the singularity
problem commonly associated with this model), an extended Krieger model,
accommodating shear thinning and aggregation effects \cite{Hund2017}, a
Casson model with haematocrit dependence following Merril et
al.~\cite{Merrill1963,Das1998}, and a modified Carreau type model,
proposed by Yeleswarapu \cite{Yeleswarapu1998}. All model parameters
have been fitted to the experimental data of Brooks \cite{Brooks1970}
(Figure \ref{fig:viscosityModels}).

    \hypertarget{krieger-dougherty-model}{%
\subsubsection{Krieger-Dougherty Model}\label{krieger-dougherty-model}}

The traditional Krieger-Dougherty model \cite{Krieger1959} was developed
to describe the rheology of high volume ratio suspensions of rigid
spherical particles. Rigid, spherical particles do not exhibit
shear-thinning behaviour, so the Krieger-Dougherty model is only
dependent on the haematocrit concentration \(\phi\). It shows a
singularity for \(\phi = \phi^*\), where \(\phi^*\) is the haematocrit
concentration for which the suspension does stop to behave like a fluid.
For rigid spheres \(\phi^* = 0.68\) \cite{Krieger1959}, while for blood
it can go up to \(\phi^* = 0.98\), which is ususally attributed to the
deformability of the erythrocytes \cite{Hund2017}.

\begin{equation} 
\mu = \mu_P \left( 1 - \frac{\phi}{\phi^*} \right)^{-n}.
\label{eq:KD}
\end{equation}

The parameter \(n = k \phi^*\) is often set to \(n=2\), but more
commonly to the high shear limit of \(n=1.82\) for \(\phi^* = 0.68\)
\cite{Quemada1978,Papir1970}, which is also the value used in this work
to allow comparison with the results from Phillips and others
\cite{Phillips1992,Mansour2010,Chebbi2018}. \(\mu_P\) is the Newtonian
viscosity of the liquid phase (plasma).

In this study the Krieger-Dougherty model is not used as for modelling
blood viscosity but as a reference model for verification and
validation.

    \hypertarget{quemada-model}{%
\subsubsection{Quemada Model}\label{quemada-model}}

The Quemada model is based on ``optimisation of viscous dissipation''
\cite{Quemada1977}. In its original form it is formulated as a
Newtonian, concentration dependent viscosity:

\begin{equation}
\mu = \mu_P \left( 1 - k \phi \right)^{-2},
\end{equation}

with \(k\) being related to the packing concentration and (for the high
shear limit) given as: \(k = 2/\phi^*\). In this form it is closely
related to the Krieger-Dougherty model (eq. \ref{eq:KD}).

In its non-Newtonian form \(k\) is expressed as
\cite{Quemada1978,Quemada1978a}:

\begin{equation}
k = \frac{k_0 + k_\infty \sqrt{\dot{\gamma}/\dot{\gamma}_c}}{1 + \sqrt{\dot{\gamma}/\dot{\gamma}_c}},
\end{equation}

where \(k_0\) and \(k_\infty\) are the intrinsic viscosities at zero and
infinite shear, respectively, and \(\dot{\gamma}_c\) is a critical shear
rate.

The shear rate magnitude \(\dot{\gamma}\) is defined as

\begin{equation}
\dot{\gamma} := \sqrt{2 \bf{D} : \bf{D}},
\end{equation}

with \(D\), the symmetric part of the velocity gradient tensor.

Different parameter fits have been proposed for
\(k_0, k_\infty, \dot{\gamma}_c\). Cokelet
\cite{Cokelet1963,Merrill1963} proposed:

\begin{eqnarray}
k_0 &=& \exp({a_0 + a_1 \phi + a_2 \phi^2 + a_3 \phi^3}) \\
k_\infty &=& \exp({b_0 + b_1 \phi + b_2 \phi^2 + b_3 \phi^3}) \\
\dot{\gamma}_c &=& \exp({c_0 + c_1 \phi + c_2 \phi^2 + c_3 \phi^3}).
\end{eqnarray}

\begin{figure}
            \begin{center}\adjustimage{max size={\linewidth}{0.7\paperheight}}{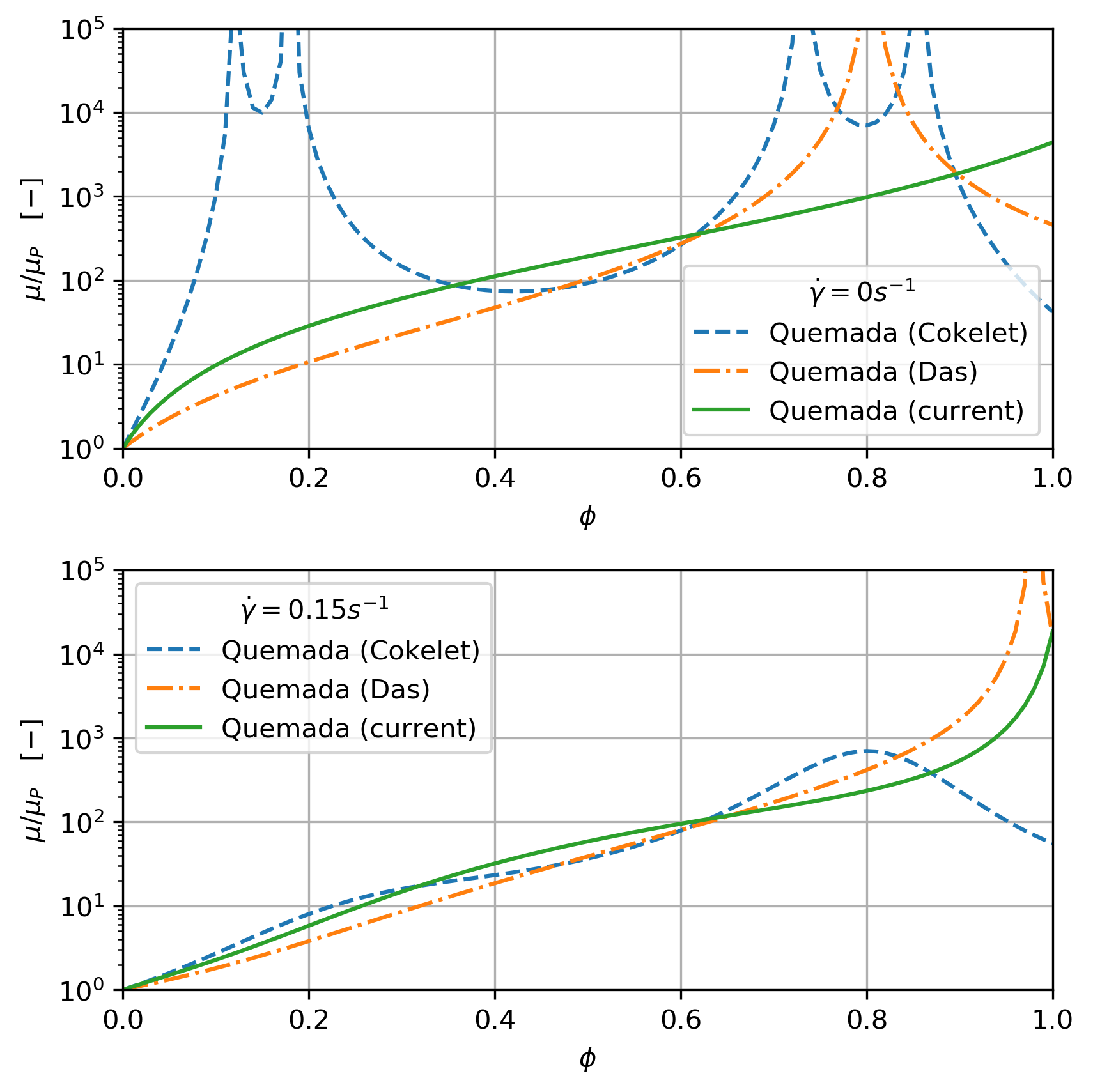}\end{center}
            \caption{Comparison of Quemada parameterisation for zero and low ($0.15 \mathrm{s}^{-1}$) shear rate. The classic Cokelet parameter set shows singularities at $12.2\%,\ 18\%,\ 73.1\%,\ 85.6\%$, the modified parameterisation by Das improves on this, but still shows a singularity for $80.4\%$ haematocrit. The current parameter set removes the singularity and shows monotonous behaviour.}
            \label{fig:QuemadaParameterComp}
        \end{figure}

Das \cite{Das1998} noted that Cokelet's parameter set causes the
viscosity to be non-monotonous over haematocrit concentration for low
shear, and exhibits singularities for zero shear. Das changed the
parameter fit for \(k_0\) to

\begin{equation}
k_0 = a_0 + \frac{2}{a_1 + \phi},
\end{equation}

which results in a monotonous behaviour for low shear (the lowest shear
measured in the Brooks dataset is around
\(\dot{\gamma} = 0.15\ \mathrm{s^{-1}}\)), but still shows a singularity
for \(\phi = 80.4\%\). While this is outside the haematocrit values
typically encountered in clinical practice, it can still pose a problem
if cell migration is taken into account, which will concentrate cells in
the core region. In order to overcome this problem, a new parameter set,
based on Das's formulation, is derived in this work, which does not show
a singularity. Figure \ref{fig:QuemadaParameterComp} shows viscosity
over shear rate for low shear rate
(\(\dot{\gamma} = 0.15\ \mathrm{s^{-1}}\)) and zero shear rate. While
all the curves show a good fit with the data, the new parameter set does
show monotonous behaviour throughout and no singularity below the
critical haematocrit.

    \hypertarget{modified-5-parameter-krieger-model}{%
\subsubsection{Modified 5 parameter Krieger
Model}\label{modified-5-parameter-krieger-model}}

Hund et al.~\cite{Hund2017} proposed and developed a quasi-mechanistic
extension to the Krieger-Dougherty model.

Starting from the traditional formulation of the Krieger-Dougherty
model:

\begin{equation} 
\mu = \mu_P \left( 1 - \frac{\phi}{\phi^*} \right)^{-n},  \label{eq:MKM5-KD}
\end{equation}

describing the haematocrit dependence, the shear-thinning behaviour is
introduced by a variable exponent \(n\):

\begin{equation}
n = n_\infty + \begin{cases}      0, \ \phi < \phi_{st} \\
                                  n_{st}, \phi > \phi_{st} , 
               \end{cases}
\end{equation}

where \(\phi_{st}\) is the threshold haematocrit concentration below
which no shear-thinning is observed. Based on Brooks \cite{Brooks1970},
this threshold is around \(\phi = 0.15\), and \(n_\infty\) is modelled
using a exponential dependency on \(\phi\):

\begin{equation}
n_\infty = a + b\ exp(-c\ \phi).
\end{equation}

Hund's \cite{Hund2017} shear-thinning exponent \(n_{st}\) comprises
contributions of red blood cell aggregation and deformability:

\begin{equation}
n_{st} = n_{agg} + n_{def},
\end{equation}

where each component is described by a power law:

\begin{equation}
n_{agg/def} = \beta_{agg/def}\ {\gamma'_{agg/def}}^{-\nu_{agg/def}},
\end{equation}

with the empirical coefficient \(\beta\) and \(\nu\), and the
non-dimensional shear rate
\(\gamma' = 1 + (\lambda \dot{\gamma})^{\nu_g}\), as defined by Carreau
and Yasuda \cite{Sequeira2007}, with a time constant \(\lambda\), and
\(\nu_g = 2\). This formulation ensures finite \(n_{st}\) at zero shear.

In the 5-component form the aggregation and deformation influences on
the shear-thinning exponent are combined into a single power law, due to
the limited data on these effects:

\begin{equation}
n_{st} = \beta \gamma'^{-\nu}.
\end{equation}

The model proposed by Hund et al.~allows for inclusion of the influence
of large molecule concentration (proteins polysacharides, lipids), as
well as fibrinogen, and temperature on the constitutive model. Due to a
lack of data these are not included in the 5-parameter model.

    \hypertarget{yeleswarapu-wu-model}{%
\subsubsection{Yeleswarapu-Wu Model}\label{yeleswarapu-wu-model}}

This model is based on a visco-elastic Oldroyd-B model developed by
Yeleswarapu et al.~\cite{Yeleswarapu1998,Wu2015}. In this study the
visco-elastic effects are neglected, only the shear-thinning behaviour
and haematocrit dependency are implemented. The shear-thinning behaviour
follows a modified Carraeu-type model based on a mixture model by Jung
et. al \cite{Jung2008}.

The model is based on a mixture model and thus the viscosity is decribed
as a function of plasma viscosity \(\mu_{P}\) and red blood cell
viscosity \(\mu_{rbc}\) \cite{Wu2015}:

\begin{equation}
\mu_{mix} = (1-\phi)\mu_{P} + \phi\mu_{rbc},
\end{equation}

where the red blood cell viscosity is described as:

\begin{equation}
 \mu_\infty(\phi) + (\mu_0(\phi) - \mu_\infty(\phi)) \frac{1 + ln(1 + k \dot{\gamma}))}{1 + k \dot{\gamma}},
\end{equation}

where, in this implementation, k is a constant model parameter, and
\(\mu_0\) and \(\mu_\infty\) are modelled as third order polynomials of
\(\phi\):

\begin{eqnarray}
    \mu_0 & = & a_1 \phi + a_2 \phi^2 + a_3 \phi^3 \\
    \mu_\infty & = & b_1 \phi + b_2 \phi^2 + b_3 \phi^3
\end{eqnarray}

    \hypertarget{casson-merrill-model}{%
\subsubsection{Casson-Merrill Model}\label{casson-merrill-model}}

The Casson model \cite{Casson1959} is a classical non-Newtonian model in
which the viscosity is modelled as:

\begin{equation}
\mu = \left( \sqrt{\mu_\infty} + \sqrt{\frac{\tau_0}{\dot{\gamma}}} \right)^2,
\end{equation}

where \(\mu_\infty\) is the Casson viscosity (asymptote at high shear
rate) and \(\tau_0\) is the yield stress. The yield effect means that
this model has a singularity at zero shear, leading to infinite
viscosity. While there is an argument that blood does exhibit yield at
slow time scales and low shear, this effect will typically make this
type of model unsuited for numerical simulation within a generalised
Newtonian approach with a local effective viscosity due to numerical
instability.

For blood, Merill et al.~gave the expressions for \(\mu_\infty\) and
\(\tau_0\) as \cite{Merrill1963,Das1998}

\begin{eqnarray}
\mu_\infty & = & \left( \frac{\mu_{pl}}{(1-\phi)^\alpha}\right) \\
\tau_0     & = & \beta^2 \left[ \left( \frac{1}{1-\phi} \right)^{\alpha/2} - 1 \right]^2,
\end{eqnarray}

with the fitting parameters \(\alpha\) and \(\beta\).

    \hypertarget{characteristics-of-rheology-models}{%
\subsubsection{Characteristics of Rheology
Models}\label{characteristics-of-rheology-models}}

All viscosity model parameters were fitted to experimental data for
varying levels of haematocrit in ADC plasma reported by Brooks
\cite{Brooks1970}. While this data is for steady state shear only, it is
still considered on of the best datasets for blood rheology data and is
used in the majority of work on blood rheology. The parameters were
fitted using a Levenberg-Marquardt least squares fit, implemented in
Scientific Python (scipy), using the MINPACK library. Table
\ref{table:viscParams} shows the parameter sets for the different
models, figure \ref{fig:viscosityModels} shows the comparison of model
results and experimental data. All models show a good fit to the
experimental data in the range were experimental data is available
(\(\dot{\gamma} > 0.15\ \mathrm{s^{-1}}\)), while the behaviour for low
shear stress varies between the models. The Casson model shows a
singularity for zero shear (yield stress behaviour), while the other
models all have finite viscosity for zero shear. However, the values at
low shear vary widely. For \(\dot{\gamma} = 10^{-2}\ \mathrm{s^{-1}}\),
the range of relative viscosity is between
\(\mu/\mu_{P} = 71.4\ \text{to}\ 936\). This variation will heavily
influence the behaviour at low shear rate, e.g.~on the axis of the flow.

Figure \ref{fig:QuemadaParameterComp} shows the parameter fit for the
Quemada model, where the classical Cokelet fit \cite{Cokelet1963}
exhibits singularities at \(12.2\%\), \(18\%\), \(73.1\%\), and
\(85.6\%\) for zero shear. The Das variation \cite{Das1998} improves on
this, but the original parameter set by Das still shows a singularity
for \(80.4\%\) haematocrit. The new parameter fit performed in this
study removes the singularities completely and shows monotonous
behaviour for the whole range of haematocrit concentrations and shear
rates.

    \begin{table}
    \begin{tabular}[]{@{}llll@{}}
\toprule
Quemada & MKM5 & Yeleswarapu & Casson\tabularnewline
\midrule
- & - & - & -\tabularnewline
a0: 0.06108 & - & a1: -0.02779 & -\tabularnewline
a1: 0.04777 & - & a2: 1.012 & -\tabularnewline
- & - & a3: -0.636 & -\tabularnewline
b0: 1.803 & b: 8.781 & b1: 0.0749 & \(\alpha\): 1.694\tabularnewline
b1: -3.68 & c: 2.824 & b2: -0.1911 & \(\beta\): 0.01197\tabularnewline
b2: 2.608 & \(\beta\): 16.44 & b3: 0.1624 & -\tabularnewline
b3: -0.001667 & \(\lambda\): 1296 & - & -\tabularnewline
- & - & k: 8.001 & -\tabularnewline
c0: -7.021 & \(\nu\): 0.1427 & - & -\tabularnewline
c1: 34.45 & - & - & -\tabularnewline
c2: -39.94 & - & - & -\tabularnewline
c3: 14.09 & - & - & -\tabularnewline
\bottomrule
\end{tabular}
\caption{Viscosity model parameters. Levenberg-Marquardt least squares fit (scipy, MINPACK), to Brooks's data (all viscosities calculated in $\mathrm{Pa\ s}$), $\mu_P = 1.23 \cdot 10^{-3}\ \mathrm{Pa\ s}$.}
\label{table:viscParams}

\end{table}
    \hypertarget{implementation}{%
\subsection{Implementation}\label{implementation}}

The model was implemented using the
\emph{Field Operation And Manipulation} (FOAM) framework (REF). FOAM, or
OpenFOAM, is an open source library which allows easy implementation of
Finite Volume Method (FVM) solvers.

The fundamental equations for mass and momentum conservation were
implemented using the SIMPLE (Semi-Implicit Method for Pressure-Linked
Equations) \cite{Patankar1972} method for steady state, and the PISO
(Pressure-Implicit with Splitting of Operators) \cite{Issa1986} and
PIMPLE (combining PISO and SIMPLE) methods for transient simulations.

Discretision is typically second order in space and time. The code
supports all discretisation methods that are supported in the FOAM
library (currently foam-extend 4.0 and OpenFOAM 1912).

The haematocrit transport equation \ref{eq:phiTransport} is implemented
as a scalar transport equation, solved outside of the SIMPLE loop. The
Laplacians in \(\phi\) are implemented implicitely (\emph{fvm}) as
diffusion terms, while the source terms in \(\dot{\gamma}\) and \(\mu\)
are calculated explicitely (\emph{fvc}).

For steady state (SIMPLE) and transient cases with the PIMPLE algorithm,
underrelaxation is required, typically the underrelaxation factor that
is required can be estimated from the order of magnitude of the ratio
between collision radius. Stable simulation has been achieved for
relaxation factors of \(0.1 \log(O(a/R))\), e.g.~a radius
\(R=50\ \mathrm{\upmu m}\) and collision radius of
\(3.5\ \mathrm{\upmu m}\) will require an underrelaxation factor of
\(\approx 0.1\) with no underrelaxation for the final iteration. The
PISO algorithm does not use underrelaxation and requires a time step to
be estimated from the Courant number (\(Co < 1\)) for \(O(a/R) > 1\),
and a smaller time step calculated based on a Courant number scaled with
the migration velocity.

The discretisation schemes used in the calculations presented in this
paper are: second order Euler \emph{backward} in time and second order
(\emph{Gauss linear}, and \emph{Gauss linear upwind} for advective
terms) in space, gradients are approximated using the
\emph{least squares} theme.

Rheology models are implemented as quasi-Newtonian, with calculation of
local cell viscosity based on the shear rate and haematocrit value in
the cell from the previous iteration/time step. The new rheology models
that are implemented at the time of writing are the standard
Krieger-Dougherty, the modified 5-parameter Krieger, the Yeleswarapu-Wu,
and the Quemada model.

    \hypertarget{results}{%
\section{Results}\label{results}}

All results shown in this paper are for fully developed pipe flow, with
periodic boundary conditions between outlet and inlet, with prescribed
average velocity. The radius of the pipe varies between \(50 \mu m\) and
\(5 mm\), to represent typical vessel diameters. The pipe length is two
diameters.

\hypertarget{verification-and-influence-of-mesh-type}{%
\subsection{Verification and influence of mesh
type}\label{verification-and-influence-of-mesh-type}}

The verification case for the implementation is a pipe of radius
\(50\ \mathrm{\upmu m}\), average velocity
\(V=0.0065\ \mathrm{m\ s^{-1}}\). The rheology model used in the
verification case is the Krieger-Dougherty model to allow comparison to
the analytical solution \cite{Phillips1992} (no analytical solution
available for the non-linear terms in the shear-stress and concentration
dependent models). Model parameters for the Krieger-Dougherty model are
\(K_c = 0.41\), \(K_{\mu} = 0.62\), \(\phi^* = 0.68\), and \(n = 1.82\).

\begin{figure}
            \begin{center}\adjustimage{max size={\linewidth}{0.7\paperheight}}{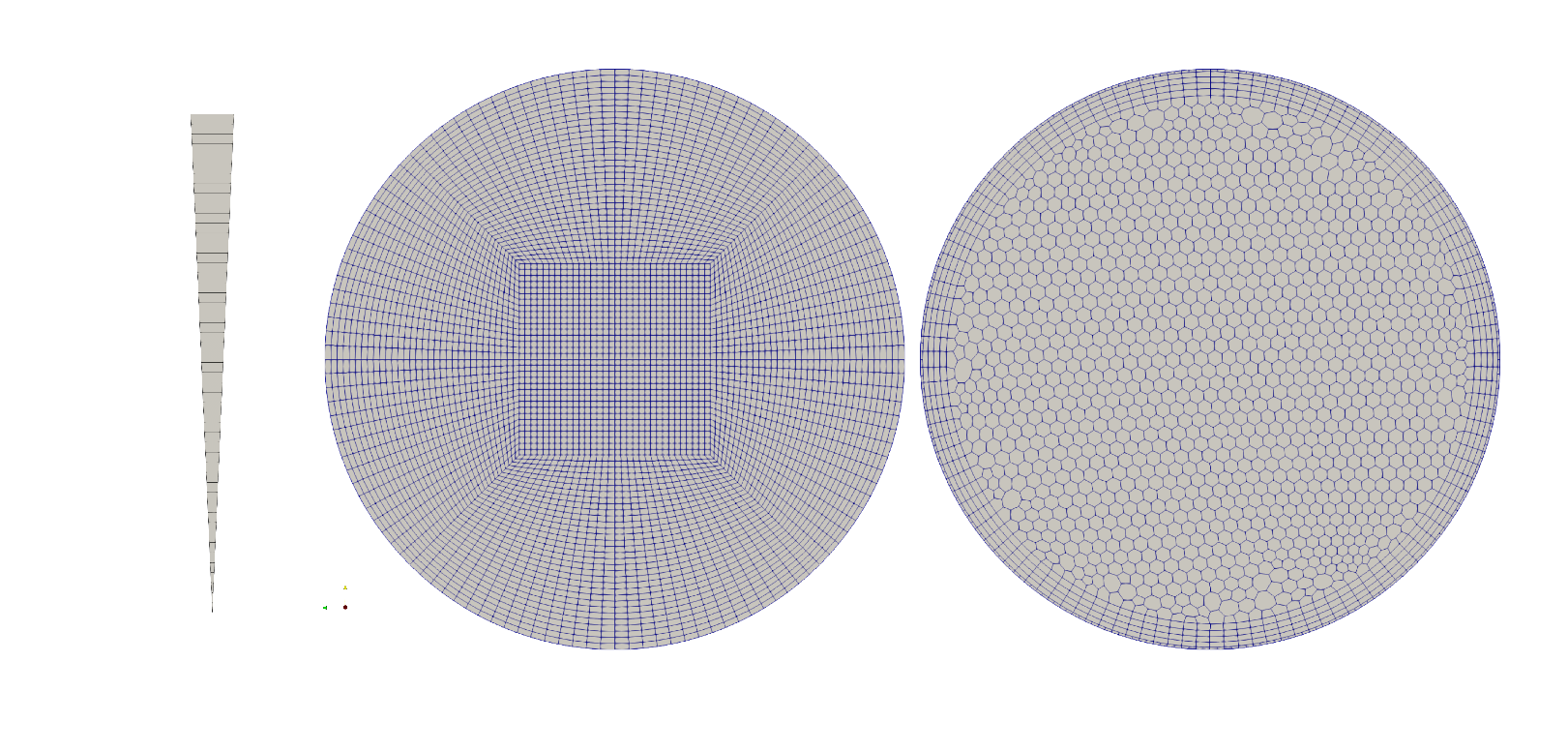}\end{center}
            \caption{Mesh topology for the verification of the model: axisymmetric wedge, 50 cells radial; hexahedral, block-structured, 50 cells radial; polyhedral with boundary layer extrusion, 60 cells diameter.}
            \label{fig:meshes}
        \end{figure}

The simulation was performed for different meshes, Figure
\ref{fig:meshes}, (a) an axisymmetric (2D) wedge with 50 cells
resolution in radial direction, (b) a hexahedral, block structured mesh
- 50 cells radial, and (c) a poyhedral mesh with boundary layer
inflation with \textasciitilde{}60 cells across the diameter - this type
of mesh is common in the simulation of vascular flow in patient specific
geometries. The given resolutions were chosen based on a mesh
convergence study and realistic mesh resolutions as typically used in
vascular simulations. The migration model requires a mesh that is of
similar resolution as meshes that aim at resolving wall shear stress
(WSS) and WSS derived metrics.

\begin{figure}
            \begin{center}\adjustimage{max size={\linewidth}{0.7\paperheight}}{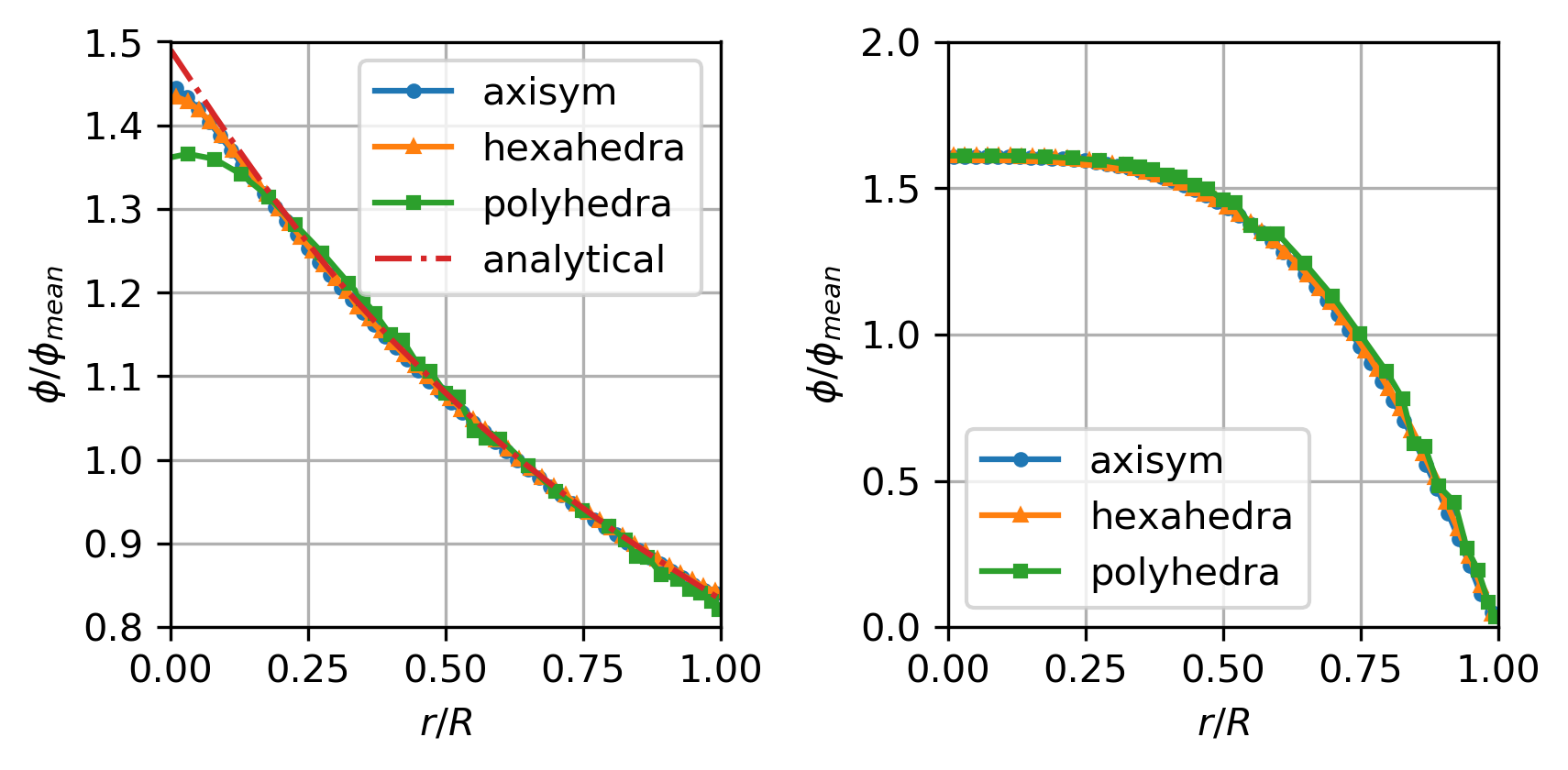}\end{center}
            \caption{Steady state particle distribution and velocity profiles for different mesh types, compared with analytical solution for particle distribution by Krieger et al. Parameters: fully developed pipe flow, $R = 50~\mathrm{\upmu m}$, $V = 0.0065~\mathrm{m~s^{-1}}$, $K_c = 0.41$, $K_\mu = 0.62$, $n = 1.82$, $\phi^* = 0.68$, Standard Krieger-Dougherty Model.}
            \label{fig:valMesh}
        \end{figure}

    Figure \ref{fig:valMesh} shows the results for the different meshes in
comparison to the analytical solution of the migration model with the
Krieger-Dougherty model. The axisymmetric two-dimensional and the
hexahedral three-dimensional meshes show excellent agreement, with only
a slight rounding of the peaked analytical solution at the axis. The
polyhedral three-dimensional mesh also shows good agreement, but the
additional numerical diffusion blunts the profile at the axis, the
concentration close to the wall is well represented.

    \hypertarget{length-and-time-scale-dependency}{%
\subsection{Length and time scale
dependency}\label{length-and-time-scale-dependency}}

    \hypertarget{wall-shear-strain-scaling}{%
\subsubsection{Wall shear strain
scaling}\label{wall-shear-strain-scaling}}

    The parabolic velocity profile for a Newtonian flow is given as:

\begin{equation}
v = -2 \, V {\left(\frac{r^{2}}{R^{2}} - 1\right)}, 
\end{equation}

where \(V\) is the average velocity.

Therefore, the velocity gradient in radial direction is:

\begin{equation}
\frac{\partial v}{\partial r} = -\frac{4 \, V r}{R^{2}}.
\end{equation}

So the gradient at the wall (\(r = R\)) scales with \(V\) and
\(R^{-1}\). The velocity is, therefore, scaled with \(R\), such that the
wall velocity gradient is constant. The Reynolds number scales with
\(R^2\). For the given values of \(R = 0.05, 0.5, 5~\mathrm{mm}\),
\(V = 0.0065,\ 0.065,\ 0.65~\mathrm{m~s^{-1}}\), the wall velocity
gradient is constant at \(\dot{\gamma}_w \approx 650~\mathrm{s^{-1}}\),
to cover the significant three decades of shear strain magnitude for
shear-thinning non-Newtonian blood models.

\begin{figure}
            \begin{center}\adjustimage{max size={\linewidth}{0.7\paperheight}}{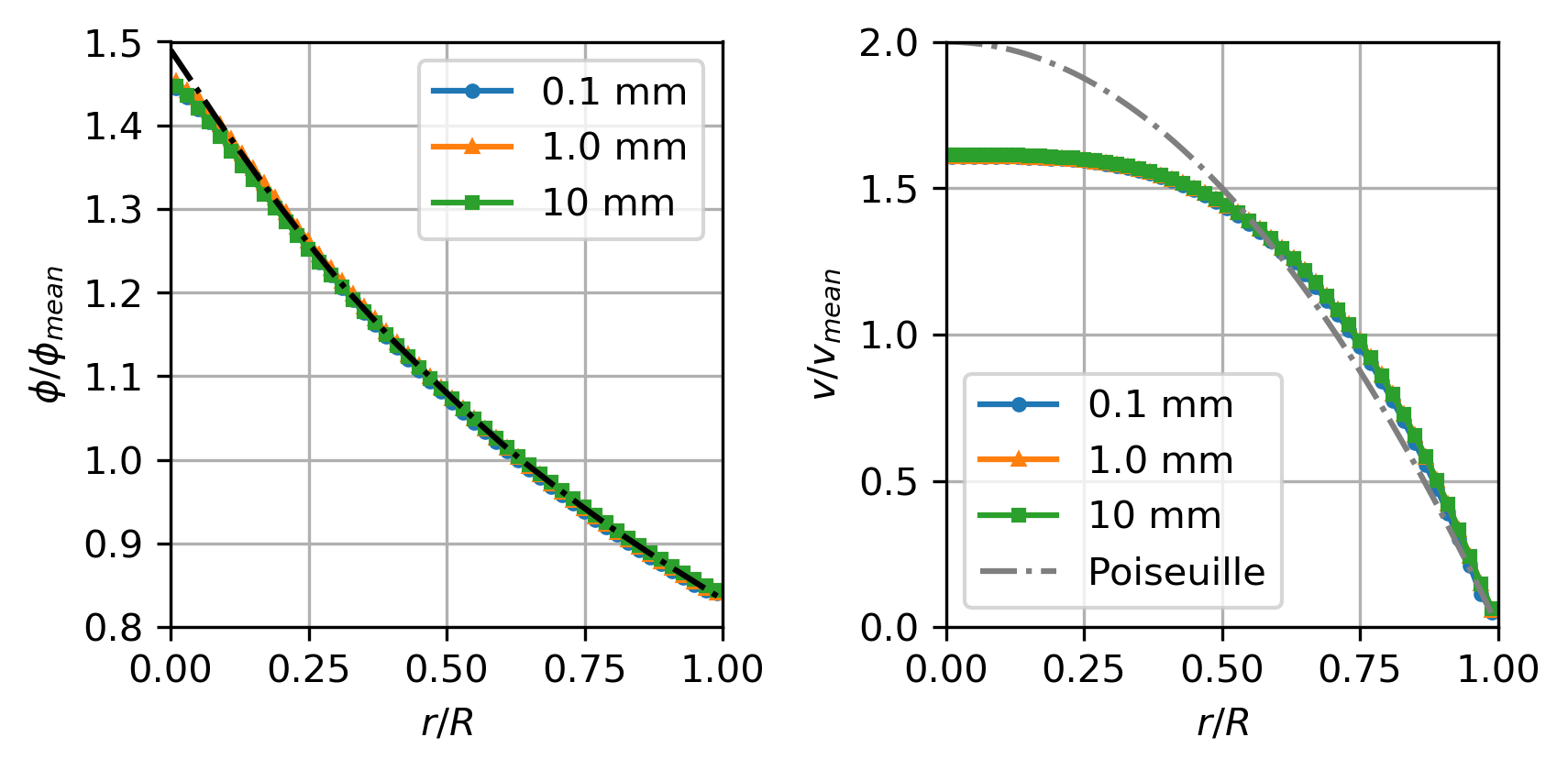}\end{center}
            \caption{Steady state particle distribution and velocity profiles for different diameters. Parameters: fully developed pipe flow, $R = 0.05,\ 0.5,\ 5~\mathrm{mm}$, $V = 0.0065,\ 0.065,\ 0.65~\mathrm{m~s^{-1}}$, $K_c = 0.41$, $K_\mu = 0.62$, $n = 1.82$, $\phi^* = 0.68$, Standard Krieger-Dougherty Model.}
            \label{fig:steady_scale}
        \end{figure}

    The steady state particle distribution profile is independent of the
length scale and the diameter ratio. It will only depend on the ratio of
\(K_c/K_\mu\). Figure \ref{fig:steady_scale} shows steady state profiles
for a range of diameters from \(0.1 - 10~\mathrm{mm}\). The
computational effort for the particle migration model, however, scales
with \(R^2/a^2\), with \(R\), the vessel radius, and \(a\), the particle
collision radius. While the small diameter \(D=0.1~\mathrm{mm}\) case is
fully converged after around \(10^4\) iterations, the
\(D=10~\mathrm{mm}\) case requires \(10^6\) iterations. This corresponds
to the diffusion timescales.

    \hypertarget{kinematic-and-particle-migration-timescales}{%
\subsubsection{Kinematic and particle migration
timescales}\label{kinematic-and-particle-migration-timescales}}

Blood flow with particle migration is governed by several different time
scales for flow kinematics and particle migration. The timescale for the
development of the velocity profile (kinematic timescale) is

\begin{equation}\label{eq:}
    \tau_k = \frac{R^2}{\nu}.
\end{equation}

The timescales for the development of the particle migration profile can
be derived from the particle migration flux diffusion terms as:

\begin{equation}\label{eq:tau_cphi}
    \tau_{c \phi} = \frac{R^2}{K_c a^2 \phi \dot{\gamma}},
\end{equation}

\begin{equation}\label{eq:tau_cgamma}
    \tau_{c \dot{\gamma}} = \frac{R^2}{K_c a^2 \phi^2},
\end{equation}

\begin{equation}\label{eq:tau_mu}
    \tau_\mu = \frac{R^2}{K_\mu a^2 \dot{\gamma} \frac{\partial{(\ln{\mu})}}{\partial \phi} }.
\end{equation}

The kinematic timescale scales with \(R^2/\nu\), while the particle
migration timescales scale with the square diameter ratio
\({R^2}/{a^2}\), where \(R\) is the pipe radius, and \(a\) is the
particle (collision) radius.

    The kinematic viscosity,
\(\nu \approx 3 \cdot 10^{-6}~\mathrm{m^{2}~s^{-1}}\), while for an
average collision radius of red blood cells of
\(a = 3.5~\mathrm{\upmu m}\), the particle migration diffusion
coefficients are of the order of
\(10^{-9} - 10^{-11} ~\mathrm{m^{2}~s^{-1}}\). This means the particle
migration happens on timescales that are three orders of magnitude
greater than the kinematic timescales.

\begin{figure}
            \begin{center}\adjustimage{max size={\linewidth}{0.7\paperheight}}{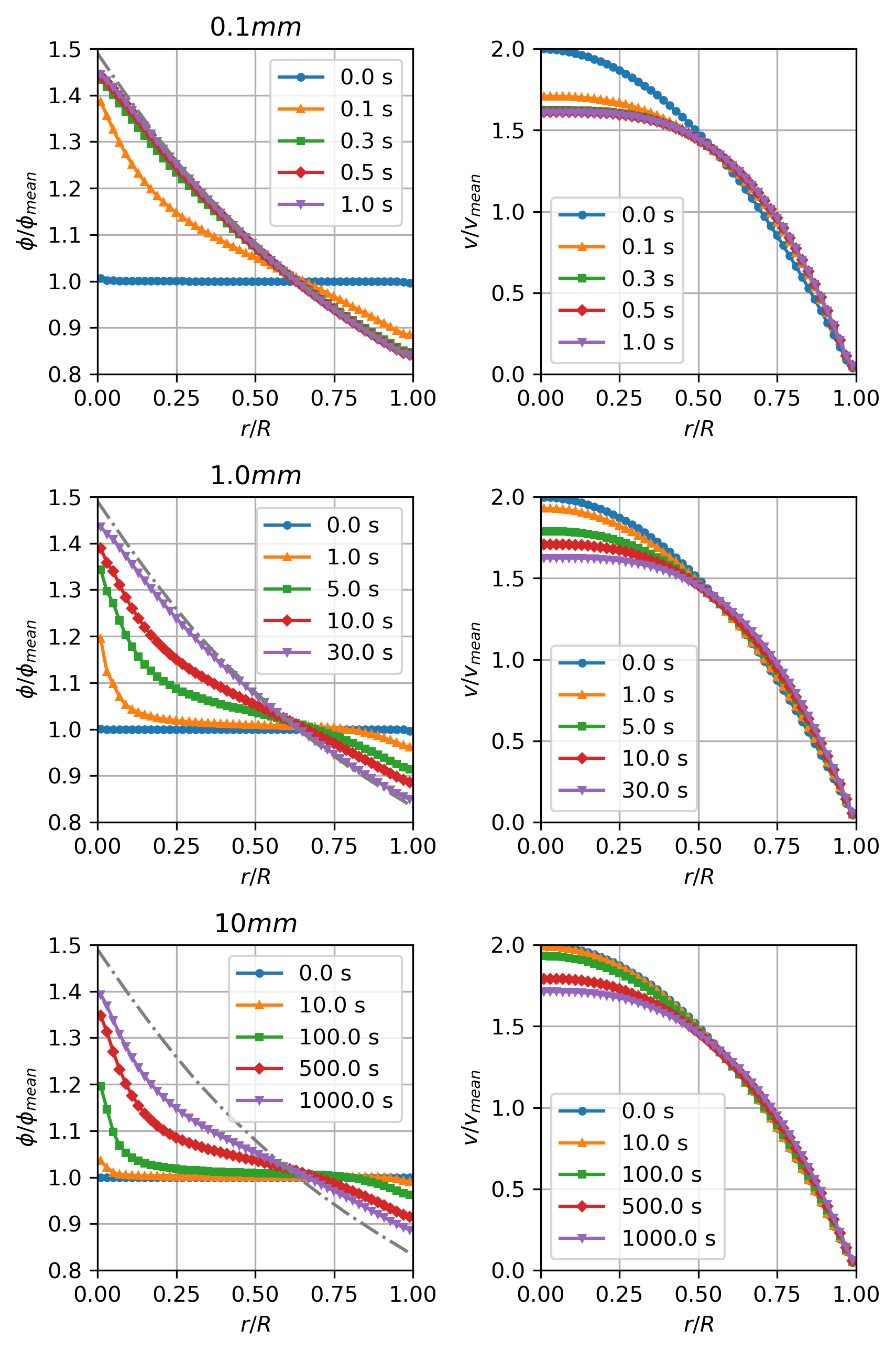}\end{center}
            \caption{Transient particle distribution and velocity profiles for different diameters. Parameters: fully developed pipe flow, $R = 0.1,\ 1.0,\ 10~\mathrm{mm}$, $V = 0.0065,\ 0.065,\ 0.65~\mathrm{m~s^{-1}}$, $K_c = 0.41$, $K_\mu = 0.62$, $n = 1.82$, $\phi^* = 0.68$, Standard Krieger-Dougherty Model.}
            \label{fig:unsteady_scale}
        \end{figure}

Figure \ref{fig:unsteady_scale} shows the temporal development of the
particle distribution and non-Newtonian velocity profile. The flows were
initialised with a fully developed parabolic velocity profile and a
uniform particle distribution of \(\phi = 0.45\) volume fraction. The
\(0.1~\mathrm{mm}\) case has reached steady state conditions within
\(0.5~\mathrm{s}\), the \(1.0~\mathrm{mm}\) case shows significant
particle migration after physiologically relevant times, while the
\(10~\mathrm{mm}\) case does show only minimal migration after
\(10~\mathrm{s}\). It can be seen that temporal scaling follows the
predicted \(R^2/a^2\) scaling factor.

    \hypertarget{variation-of-rheology-model-and-collision-parameter-ratio}{%
\subsection{Variation of rheology model and collision parameter
ratio}\label{variation-of-rheology-model-and-collision-parameter-ratio}}

As is obvious from equation \ref{eq:phiTransport}, the particle
migration is strongly dependent on the viscosity model and the balance
between collision and viscosity driven migration, as expressed in the
model parameters \(K_c\) and \(K_{\mu}\).

\begin{figure}
            \begin{center}\adjustimage{max size={\linewidth}{0.7\paperheight}}{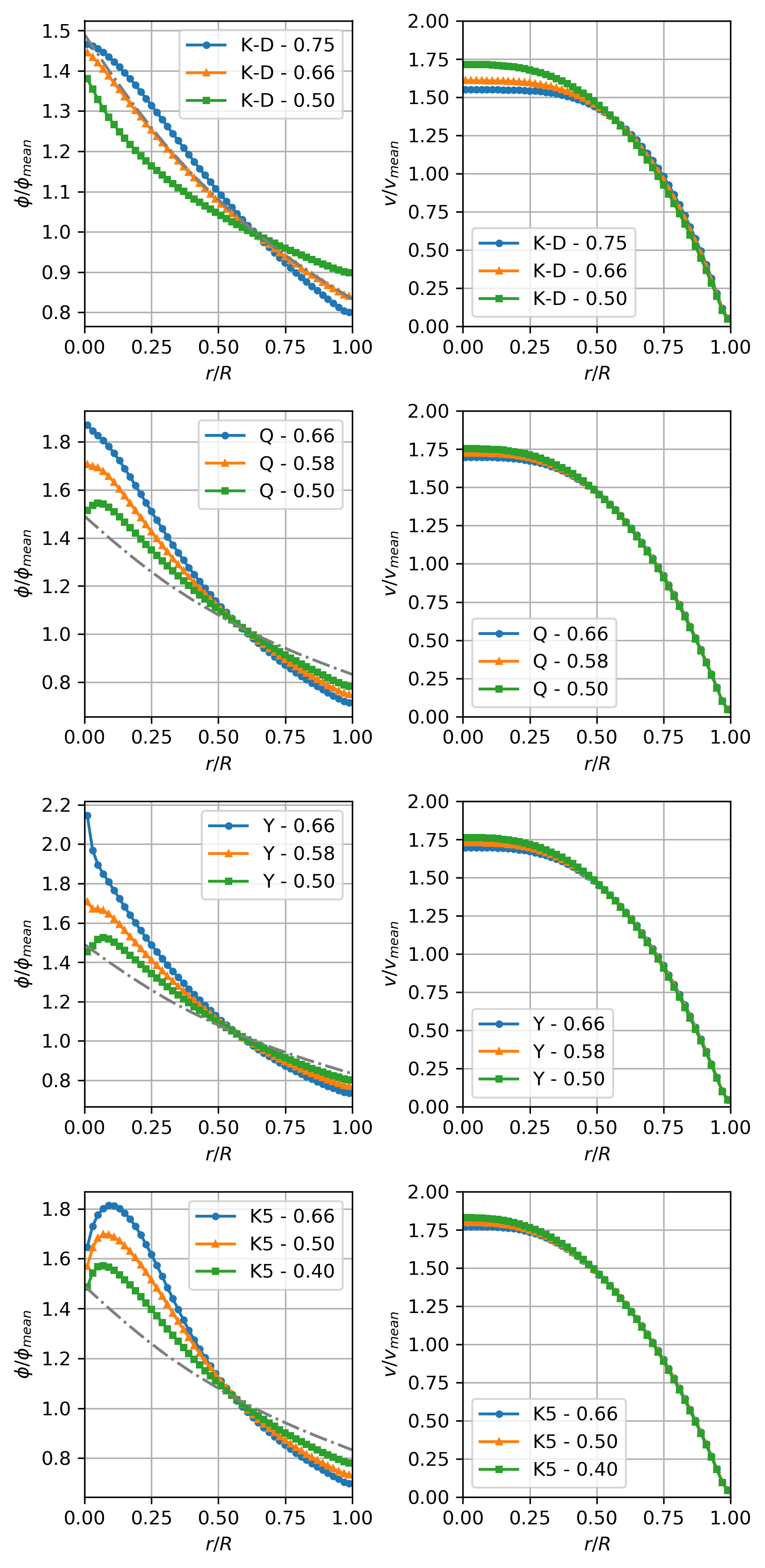}\end{center}
            \caption{Steady state particle distribution and velocity profiles for different viscosity models and collision parameter ratios ($K_c/K_\mu$). Parameters: fully developed pipe flow, $R = 0.05~\mathrm{mm}$, $V = 0.0065~\mathrm{m~s^{-1}}$, $K_c/K_\mu = 0.4 - 0.75$. Standard Krieger-Dougherty, Quemada, Yeleswarapu, and modified 5-parameter Krieger model.}
            \label{fig:steadyVisc}
        \end{figure}

While the magnitude of \(K_c\) and \(K_{\mu}\) controls the magnitude of
the migration pressures and thus the temporal response of the system,
the concentration profile only depends on the balance between collision
and viscosity driven fluxes. This balance is expressed by the ratio
between the parameters \(K_c/K_{\mu}\). Figure \ref{fig:steadyVisc}
shows the haematocrit profiles as they develop for different viscosity
models - Krieger-Dougherty (K-D), Quemada (Q), Yeleswarapu-Wu (Y),
modified 5-parameter Krieger model (K5), and varying K-ratios
\(K_c/K_{\mu} = 0.4\ \text{to}\ 0.75\).

Compared to the verification K-D case with K-ratio of \(0.66\), it can
be seen that a shift in the balance to higher influence of the collision
frequency (higher K-ratio) steepens the profile, while a lower K-ratio,
i.e.~a shift of the balance to the resistive influence of the viscosity
increase in the low shear region causes a flatter profile.

Comparing the different viscosity models clearly shows the main
difference in the core region, where the strong variation in the low
shear behaviour, discussed earlier, leads to a strong variation in the
relative viscosity gradient (last term in equation\{eq:phiTransport\}).
It is obvious that there is a need for further study and comparison with
experimental or meso-scale modelling data to find realistic parameters
for each of the potential viscosity models. Especially the modified
Krieger model (K5) shows a, most likely unrealistic, double-bump profile
at the axis.

Based on these preliminary results, the Quemada model with a K-ratio of
between \(0.5\) and \(0.6\) seems to be the most promising candidate for
a semi-mechanistic rheology model for blood.

    \hypertarget{discussion}{%
\section{Discussion}\label{discussion}}

While previous implementations \cite{Mansour2010,Chebbi2018} of this
class of model are limited by the fact that the viscosity term in
equation \ref{eq:phiTransport} is linearised in the viscosity gradient
with H, our implementation avoids this by implementing the non-linear
term directly which allows to use different viscosity models. Our
implementation also avoids the use of artificial stabilisation terms
that lead to underestimation of RBC migration \cite{Biasetti2014}.

The particle migration time scales with \((a/R)^2\), where \(a\) is the
RBC collision radius. This means that the particle migration is most
relevant for small vessels of a diameter of \(1\ mm\) or lower, where
the migration occurs on physiologically relevant timescales. For larger
vessels, minor effects caused by a synergy of particle migration and
secondary flows \cite{Biasetti2014}.

The parameters for the migration model would need to be calibrated to
experimental data. While such data is available, albeit scarce, for
rigid particles in suspension, e.g.~based on nuclear magnetic resonance
measurements of particle profiles, the authors are not aware of any such
data for soft vesicles, in particular RBCs. We therefore hope to use
meso-scale models (MCLBM) modeling the cell scale interactions to derive
integral diffusion and particle migration measures that can be used to
fit the continuum model parameters.

It has to be noted that the implementation uses the magnitude of the
shear in the particle flux formulation. As noted by Phillips
\cite{Phillips1992} this assumes an essentially one-dimensional shear
state, and isotropic response. This limits the application of the model
to flow situations where the shear tensor is aligned with the flow and
the main shear in radial direction. As with isotropic turbulence
modelling the isotropic migration model will overpredict migration
pressure in regions with high anisotropy, e.g.~stagnation points, strong
acceleration, or rotational shear. It is planned to implement an
explicit formulation for a localised, anisotropic shear and migration
pressure tensor, similar to approaches proposed by Miller
\cite{Miller2009} or Fang et al.~\cite{Fang2002}.

    \hypertarget{software}{%
\section{Software}\label{software}}

The continuum-scale haemorheology framework was implemented in
foam-extend, version 4.0/4.1, and OpenFoam, version 1912. The software
(\texttt{haemoFoam}) is freely available to interested parties on github
(\texttt{TS-CUBED/haemoFoam}). Please contact the author for testing and
developer access.

\texttt{haemoFoam} is a modelling framework for vascular flow simulation
based on FOAM, that is intended to cater for the particular requirements
of haemodynamics, in particular with respect to WSS related phenomena
like atherosclerosis. At the time of writing it includes:

\begin{itemize}
\tightlist
\item
  Haematocrit transport model, modelling the shear driven transport of
  red blood cells in direction of the shear gradient
\item
  Blood specific non-Newtonian rheology models including haematocrit
  dependency and shear thinning behaviour

  \begin{itemize}
  \tightlist
  \item
    Krieger Dougherty (non shear-thinning)
  \item
    Modified K-D \cite{Hund2017} (shear-thinning)
  \item
    Quemada
  \item
    Yeleswarapu
  \item
    Casson-Merrill
  \item
    Carreau model (not concentration dependent, Fluent implementation)
  \end{itemize}
\item
  post-processing for WSS and established WSS derived parameters:

  \begin{itemize}
  \tightlist
  \item
    TAWSS, TAWSSMag
  \item
    OSI
  \item
    transverse WSS
  \item
    Relative Residence Time
  \item
    temporal and spatial WSS gradients
  \end{itemize}
\end{itemize}

Planned future features are:

\begin{itemize}
\tightlist
\item
  Windkessel boundary conditions for outlets
\item
  viscoelastic rheology models (e.g.~Oldroyd B)
\item
  platelet transport
\item
  low density lipoprotein (LDL) transport
\item
  fluid-structure-interaction (FSI) for flexible vessel walls
\end{itemize}

    \hypertarget{references}{%
\bibliographystyle{abme} \
\bibliography{Library}}\
\end{document}